\documentclass{ws-procs9x6}            % default, citations in superscript
\usepackage[caption=false]{subfig}
\usepackage[capitalise]{cleveref}
\usepackage{xspace}
\usepackage{siunitx}
\DeclareSIUnit{\clight}{\text{\ensuremath{c}}}
\DeclareSIUnit[per-mode=symbol]\MeVcc{\MeV\per\clight\squared}
\DeclareSIUnit[per-mode=symbol]\GeVcc{\GeV\per\clight\squared}
\DeclareSIUnit[per-mode=symbol]\GeVc{\GeV\per\clight}
\DeclareSIUnit[per-mode=symbol]\GeVcsq{(GeV/\clight)^2}   

\providecommand*{\abs}[1]{}                                                                                                                                 
\renewcommand*{\abs}[1]{\ensuremath{\vert{#1}\vert}\xspace}
 
\newcommand*{\Kpipi}{\ensuremath{{K^-\pi^-\pi^+}}\xspace}      
\newcommand*{\Kpi}{\ensuremath{{K^-\pi^+}}\xspace}   
\newcommand*{\pipi}{\ensuremath{{\pi^-\pi^+}}\xspace}   
\newcommand*{\mTwoPi}{\ensuremath{m_\pipi}\xspace}

\newcommand*{\mKpipi}{\ensuremath{m_{K\pi\pi}}\xspace}  
\newcommand*{\tpr}{\ensuremath{{t'}}\xspace}

\newcommand*{\WaveK}[7]{\ensuremath{{#1}^{{#2}}\,\allowbreak{#3}^{#4}\,\allowbreak{#5}\,{#6}\,{#7}}\xspace}

\newcommand{\threeFigureSubfigureWidth}{0.333\linewidth}

\usepackage[textwidth=\marginparwidth,
textsize=tiny,
backgroundcolor=red!10,
linecolor=red,
bordercolor=red,
english]{todonotes}

\begin{document}
\title{Strange-Meson Spectroscopy at COMPASS}

\author{S. Wallner$^*$ for the COMPASS collaboration}

\address{Physics Department, Technical University of Munich,\\
Garching, 85748, Germany\\
$^*$E-mail: stefan.wallner@tum.de}

\begin{abstract}

COMPASS is a multi-purpose fixed-target experiment at CERN aimed at studying the structure and spectrum of hadrons.
It has collected the so far world's largest data set on diffractive production of the \Kpipi decay, which in principle gives access to all kaon states.
We performed an elaborate partial-wave analysis, using model-selection techniques to select the wave set based on a large systematically constructed pool of allowed partial waves.
The partial-wave decomposition reveals signals in the mass region of well-known states, such as $K_1(1270)$ and $K_1(1400)$.
%However, we observe systematic effects in some partial waves, e.g. in the $1^+0^+K^*(892)\pi S$ wave, from the limited kinematic range of our final-state particle identification.
In addition, we observe potential signals from excited states, such as $K_1(1650)$.
\end{abstract}


\bodymatter

%=============================================================================
%=============================================================================
%=============================================================================
\section{Meson Spectroscopy at COMPASS}
COMPASS is a fixed-target multi-purpose experiment located at CERN.
%\SI{190}{\GeVc} positive and negative secondary hadron beams or a tertiary muon beam are directed onto various types of targets.
%Two CEDAR detectors before the target allow to identify the beam particle.
%The forward-going final-state particles are measured by a two-stage magnetic spectrometer and identified by a ring-imaging Cherenkov detector~\cite{Abbon:2014aex}.
So far, COMPASS has studied mainly isovector resonances of the $a_J$ and $\pi_J$ families with high precision, using the dominant $\pi^-$ component of the \SI{190}{\GeVc} negative hadron beam.~\cite{Adolph2015,Akhunzyanov2018}
In this analysis, we study the strange-meson spectrum up to masses of \SI{3}{\GeVcc} using the approximately \SI{2.4}{\percent} $K^-$ fraction of the beam in diffractive scattering off a liquid-hydrogen target.
%Due to their very short lifetime, we observe the resonances only in their decays into quasi-stable final-state particles.
Our flagship channel is the production of $K^-\pi^-\pi^+$, which in principle gives access to all kaon states, i.e. $K_J$ and $K^*_J$ mesons.\footnote{Except for $J^P = 0^+$ states.}
COMPASS acquired the so far world's largest data set of about \num{720000} exclusive events for this channel.

%=============================================================================
%=============================================================================
%=============================================================================
\section{Kinematic Distributions}
\begin{figure}%
	\centering%
	\subfloat[]{\includegraphics[width=\threeFigureSubfigureWidth]{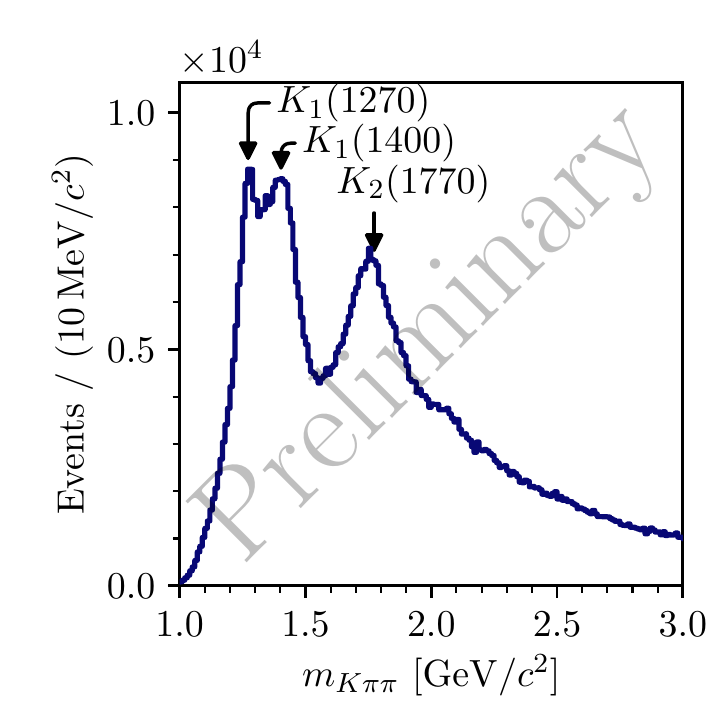}\label{fig:kin:mKpipi}}%
	\subfloat[]{\includegraphics[width=\threeFigureSubfigureWidth]{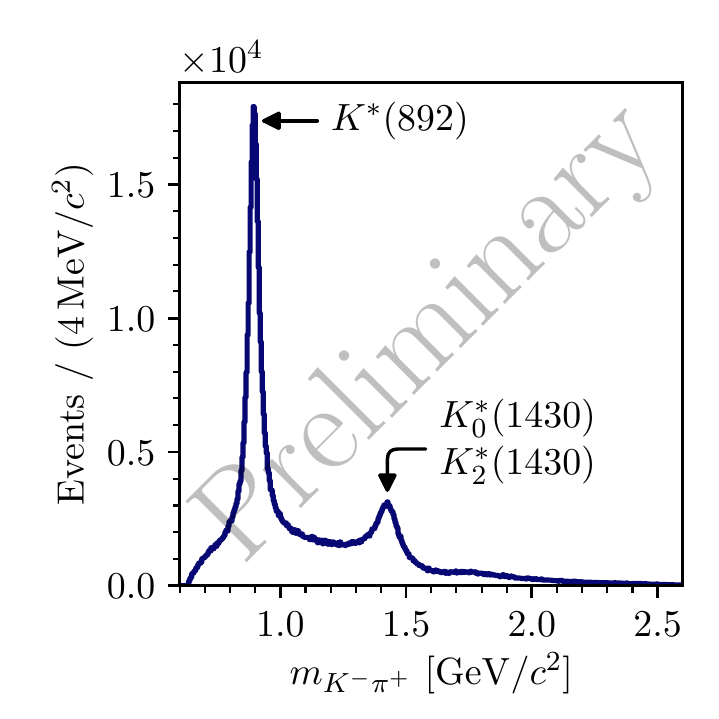}\label{fig:kin:mKpi}}%
	\subfloat[]{\includegraphics[width=\threeFigureSubfigureWidth]{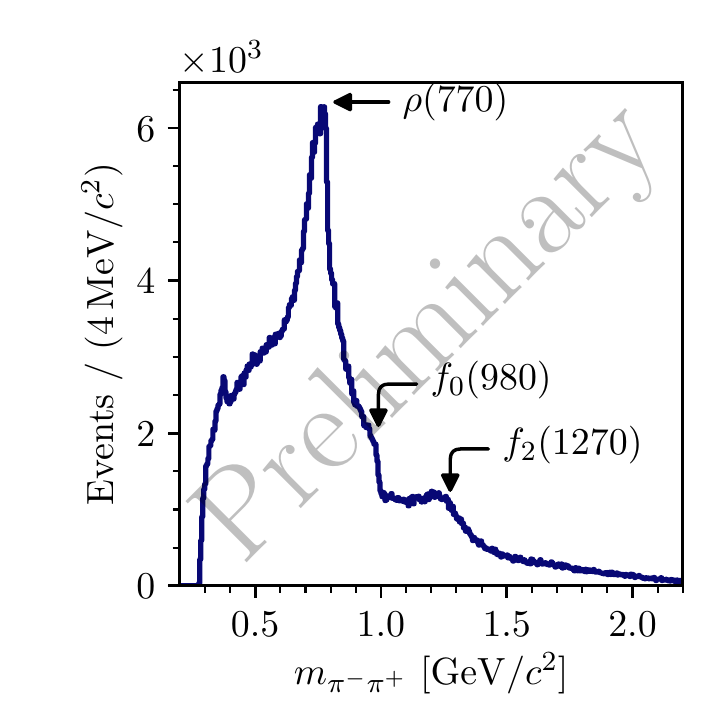}\label{fig:kin:mpipi}}%
	\caption{Invariant mass distribution of the \Kpipi system (a) and of the \Kpi (b) and \pipi (c) sub-systems. The distributions are not corrected for acceptance effects, which are significant in certain regions of the phase space.}%
	\label{fig:kin}%
\end{figure}

\Cref{fig:kin:mKpipi} shows the invariant mass spectrum of the \Kpipi final state. It exhibits structures in the \mKpipi region of well-known resonances, e.g. $K_1(1270)$ or $K_1(1400)$, which sit on top of a broad distribution.
The invariant mass distributions of the \Kpi and \pipi sub-systems are dominated by well-known two-body resonances as indicated by the labels in \cref{fig:kin:mKpi,fig:kin:mpipi}.\footnote{The small spike at about \SI{0.4}{\GeVcc} in the \mTwoPi system originates from $\phi \to K^-K^+$ decays from a small contamination of the \Kpipi data sample by $K^-K^-K^+$, where the kaons are misidentified as pions.}
This dominance of two-body resonances justifies the application of the isobar model in the partial-wave decomposition, which is discussed in the following section.

%=============================================================================
%=============================================================================
%=============================================================================
\section{Analysis Method: Partial-Wave Decomposition}
We employ the method of partial-wave analysis to decompose the data into contributions from various partial waves (see Ref.~\citenum{Adolph2015} for details).
To this end, we construct a model for the intensity distribution $\mathcal{I}(\tau)$ of the \Kpipi final state in terms of the five phase-space variables that are represented by $\tau$. Using the isobar approach, $\mathcal{I}(\tau)$ is modeled as a coherent sum of partial-wave amplitudes
\begin{equation}
\mathcal{I}(\tau; \mKpipi, \tpr) = \left|\sum\limits_{a}^{\text{waves}} {\mathcal T_a(\mKpipi, \tpr)}\, {\psi_a(\tau;\mKpipi)}\right|^2. \label{eq:intensity}
\end{equation}
These partial waves are represented by $a = \WaveK{J}{P}{M}{\varepsilon}{\zeta}{b}{L}$ and are defined by the quantum numbers of the \Kpipi system ($J^PM^\varepsilon$),\footnote{Here, $J$ is the spin of the \Kpipi state and $P$ its parity. The spin projection of $J$ along the beam axis is given by $M^\varepsilon$.} the intermediate two-body resonance $\zeta$\footnote{We consider isobar resonances in the \Kpi or \pipi sub-system.}, and the orbital angular momentum $L$ between the bachelor particle $b$ and the isobar.
Within the isobar model, the decay amplitudes $\psi_a$ can be calculated. This allows us to extract the partial-wave amplitudes $\mathcal T_a$, which determine strength and phase of each wave from the data by an unbinned maximum-likelihood fit.

In order to extract the dependence of the partial-wave amplitudes on the invariant mass \mKpipi of the \Kpipi system and on the squared four-momentum transfer \tpr between the beam particle and the target proton, the maximum-likelihood fit is performed independently in 75 narrow bins of \mKpipi and 4 bins of \tpr in the analyzed kinematic range of $1.0 < \mKpipi < \SI{3.0}{\GeVcc}$ and $0.1 < \tpr < \SI{1.0}{\GeVcsq}$.
%\footnote{By definition, \tpr is a positive quantity, as $\tpr\equiv \abs{t} - \abs{t}\sub{min}$.}
%In addition, the size of the dataset allows us to study the dependence of the partial-wave amplitudes on the squared four-momentum transfer \tpr, by binning the data in \tpr as well.
%By performing the partial-wave decomposition independently in $75$ \mKpipi bins in the range $1.0 < \mKpipi < \SI{3.0}{\GeVcc}$ and $4$ \tpr bins in the range $0.1 < \tpr < \SI{1.0}{\GeVcsq}$,\footnote{By definition, \tpr is a positive quantity, as $\tpr\equiv \abs{t} - \abs{t}\sub{min}$.} we extract simultaneously the \mKpipi and \tpr dependence of the partial-wave amplitudes from the data.

In order to construct the wave set, i.e. the partial waves that enter the sum in \cref{eq:intensity}, we apply model-selection techniques.
We systematically construct a large set of possible partial waves, called wave pool.
We restrict ourselves to $J\leq 7$, $L\leq 7$, positive naturality of the exchange particle, and twelve isobars: two $K\pi$ $S$-wave amplitudes, $K^*(892)$, $K^*(1680)$, $K^*_2(1430)$, $K^*_3(1780)$, a broad $\pi\pi$ $S$-wave amplitude, $f_0(980)$, $f_0(1500)$, $\rho(770)$, $f_2(1270)$, and $\rho_3(1690)$.\footnote{We use relativistic Breit-Wigner amplitudes for all isobar line shapes, except for the $S$-wave isobars, where we employ $K$-matrix parameterizations.}
This results in a wave pool of 596 allowed waves.
In order to select the wave set, we fit the wave pool to the data applying regularization techniques to suppress insignificant waves (see Ref.~\citenum{kaspar2019} for details).
This results in an individual wave set for each $(\mKpipi, \tpr)$ cell, which we fit again to the data without regularization terms.

%=============================================================================
%=============================================================================
%=============================================================================
\section{Selected Results of the Partial-Wave Decomposition}

%=============================================================================
%=============================================================================
\subsection{$J^P$ = $1^+$ Waves}
\begin{figure}%
	\centering%
	\subfloat[]{\includegraphics[width=\threeFigureSubfigureWidth]{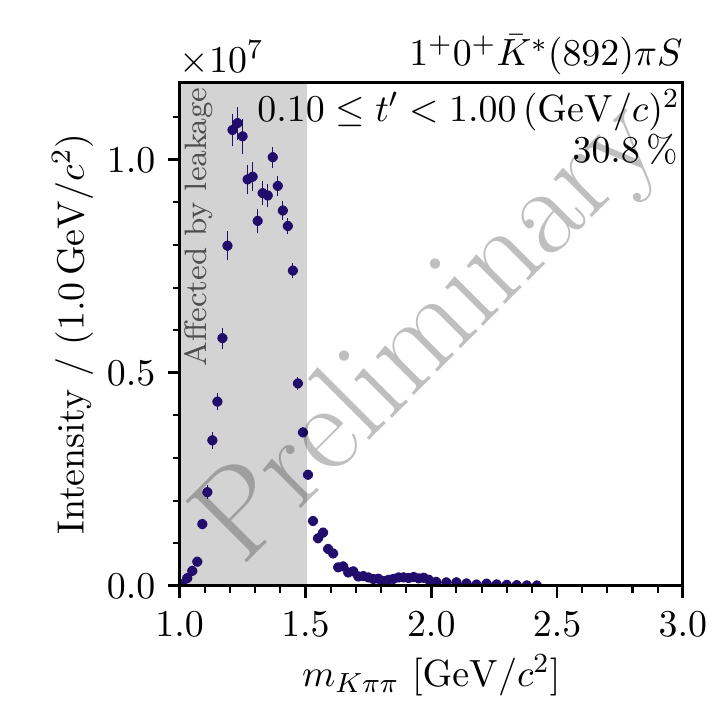}\label{fig:1+:Kst}}%
	\subfloat[]{\includegraphics[width=\threeFigureSubfigureWidth]{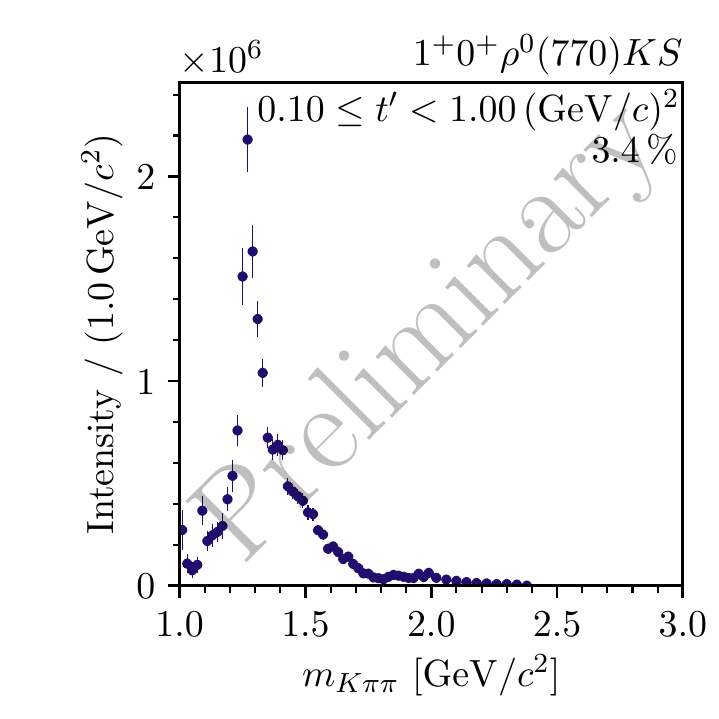}\label{fig:1+:rho}}%
	\subfloat[]{\includegraphics[width=\threeFigureSubfigureWidth]{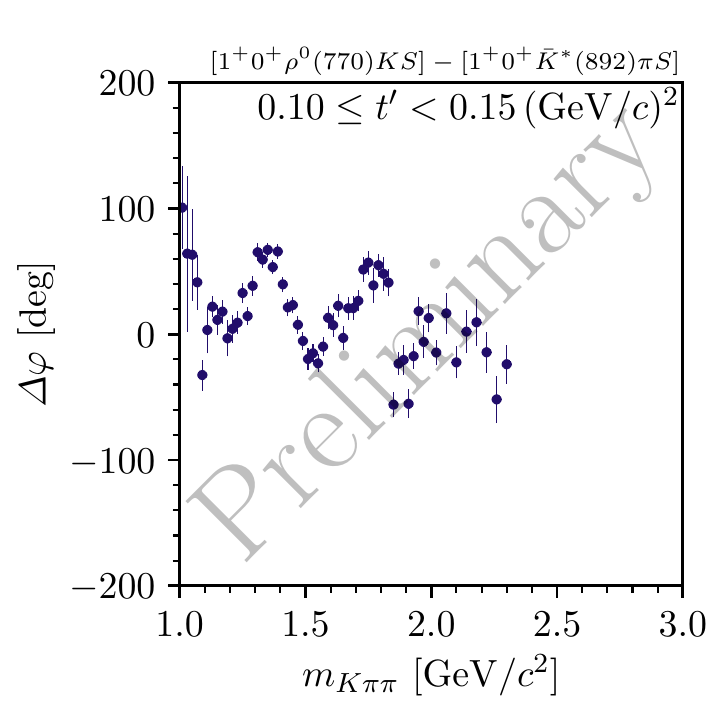}\label{fig:1+:phase}}%
	\caption{\tpr-summed intensity distributions of the $J^PM^\varepsilon = 1^+0^+$ partial waves in the $K^*(892)\pi S$ (a) and the $\rho(770)KS$ (b) decay. (c) shows the relative phase between these two partial waves in the lowest \tpr bin.}%
	\label{fig:1+}%
\end{figure}

The $1^+0^+K^*(892)\pi S$ wave is the most dominant one in our data. It accounts for about \SI{30}{\percent} of the total intensity. \cref{fig:1+:Kst} shows the \tpr-summed intensity spectrum, i.e. the intensity spectrum incoherently summed over the four \tpr bins, of the $1^+0^+K^*(892)\pi S$ wave.
Two peaks are visible in the mass regions of the $K_1(1270)$ and the $K_1(1400)$ in agreement with previous observations.~\cite{Daum1981a}
Systematic studies of our analysis revealed, that the low-mass region, indicated by the gray area in \cref{fig:1+:Kst}, of this wave and some other waves not discussed here is affected by systematic effects. These effects arise mainly from the limited kinematic range of the COMPASS final-state particle identification and are under further study. Therefore, the low-mass region of the $1^+0^+K^*(892)\pi S$ wave should be considered only on a qualitative level.

The intensity spectrum of the $1^+0^+\rho(770)KS$ wave, shown in \cref{fig:1+:rho}, exhibits only one dominant peak in the mass region of the lower-lying $K_1(1270)$. We observe a bump in its high-mass shoulder at about \SI{1.6}{\GeVcc}. The PDG~\cite{Tanabashi2018} lists the $K_1(1650)$ resonance in this mass region, which needs further clarification.

\Cref{fig:1+:phase} shows the relative phase between the $1^+0^+\rho(770)KS$ and the $1^+0^+K^*(892)\pi S$ wave. It exhibits rich structures.
First, the phase rises in the $K_1(1270)$ mass region, which can be explained by the dominance of the $K_1(1270)$ in the $\rho(770)KS$ decay.\footnote{The phase motion produced by the  $K_1(1270)$ resonance in the $K^*(892)\pi S$ decay is potentially smeared out by the low-mass tail of the $K_1(1400)$.}
Then, the relative phase drops in the mass region of the $K_1(1400)$. This can be explained by the strong contribution of the $K_1(1400)$ to the $K^*(892)\pi S$ decay, which enters with a minus sign in the relative phase.
Finally, the relative phase rises again at about \SI{1.7}{\GeVcc}. This could be a potential indication of the $K_1(1650)$ resonance in the $\rho(770)K S$ decay.

%=============================================================================
%=============================================================================
%\subsection{$J^P$ = $2^+$ Waves}
%
%\begin{figure}%
%	\centering%
%	\subfloat[]{\includegraphics[width=\threeFigureSubfigureWidth]{pictures/2+_1+_Kstarbar_892_0_pi-_D__tPrime_summed.pdf}\label{fig:2+:Kst}}%
%	\subfloat[]{\includegraphics[width=\threeFigureSubfigureWidth]{pictures/2+_1+_rho_770_0_K-_D__tPrime_summed.pdf}\label{fig:2+:rho}}%
%	\subfloat[]{\includegraphics[width=\threeFigureSubfigureWidth]{pictures/2+_1+_rho_770_0_K-_D__1+_0+_Kstarbar_892_0_pi-_S__tPrime_0-150_0-240.pdf}\label{fig:2+:phase}}%
%	\caption{\tpr-summed intensity distribution of the $J^PM^\varepsilon = 1^+0^+$ partial waves in the $K^*(892)\pi^-S$ (a) and the $\rho(770)K^-S$ (b) decay. (c) shows the relative phase between these two partial waves in the first \tpr bin.}%
%	\label{fig:2+}%
%\end{figure}

%=============================================================================
%=============================================================================
\subsection{$J^P$ = $2^-$ Waves}

\begin{figure}%
	\centering%
	\subfloat[]{\includegraphics[width=\threeFigureSubfigureWidth]{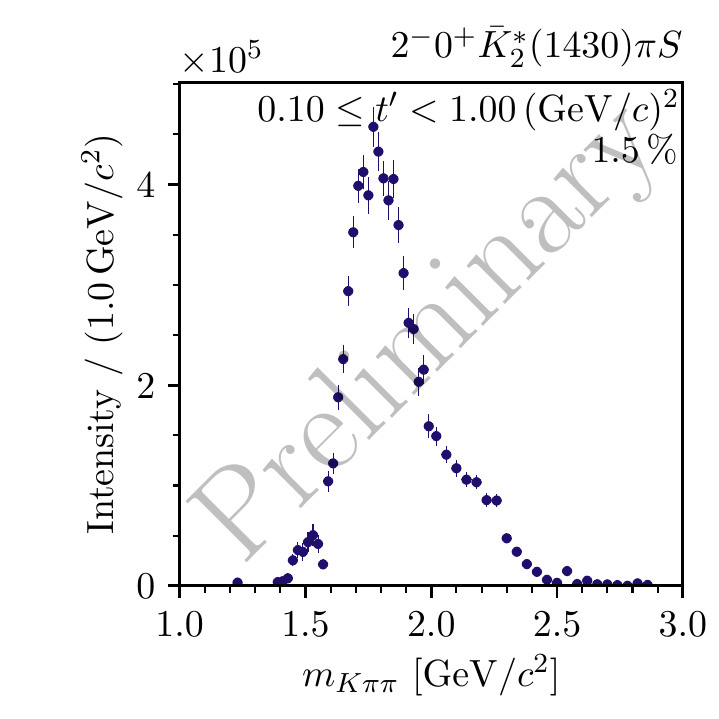}\label{fig:2-:K2st}}%
	\subfloat[]{\includegraphics[width=\threeFigureSubfigureWidth]{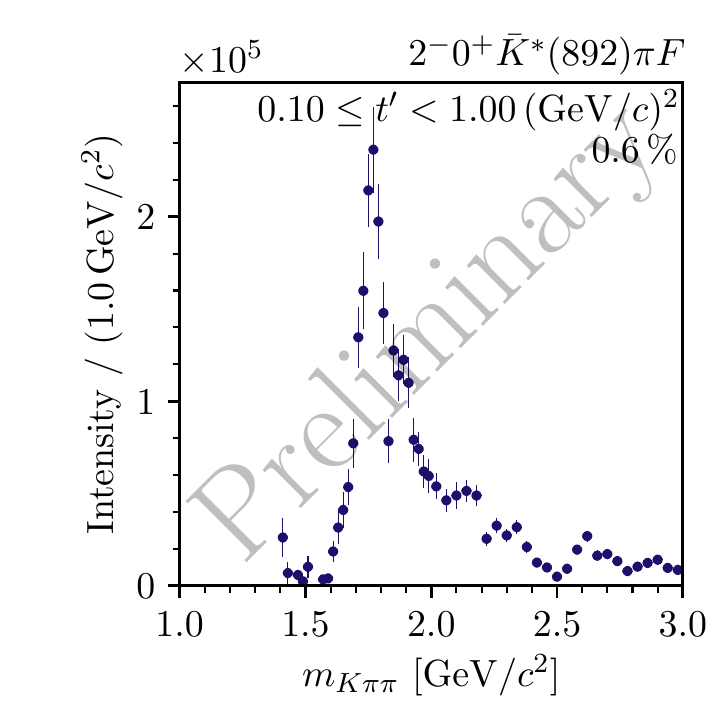}\label{fig:2-:Kst}}%
	\subfloat[]{\includegraphics[width=\threeFigureSubfigureWidth]{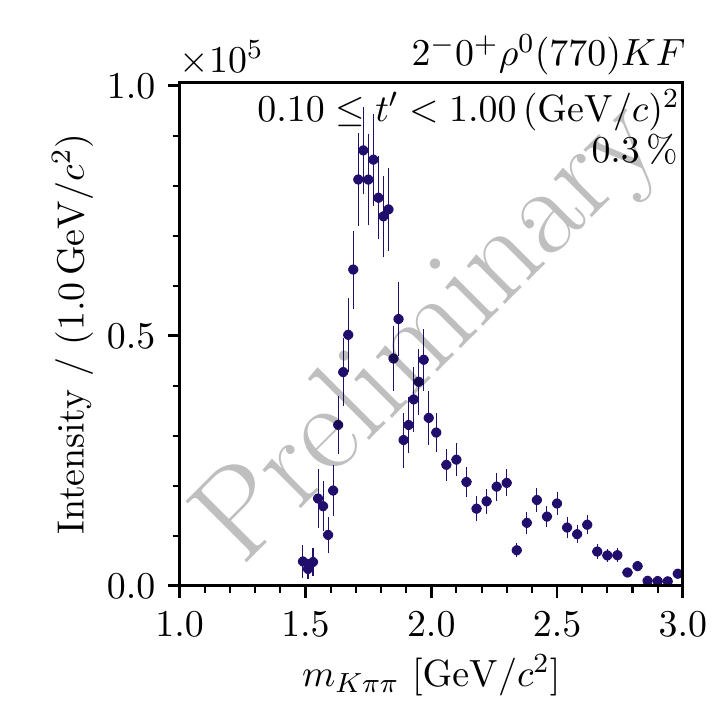}\label{fig:2-:rho}}%
	\caption{\tpr-summed intensity distributions of the $J^PM^\varepsilon = 2^-0^+$ partial waves in the $K^*_2(1430)\pi S$ (a), $K^*(892)\pi F$ (b), and the $\rho(770)FS$ (c) decay.}%
	\label{fig:2-}%
\end{figure}

The $2^-$ waves show a large variety of different structures in various decay modes.
The most dominant signal is in the $2^-0^+ K^*_2(1430) \pi S$ wave shown in \cref{fig:2-:K2st}, which accounts for about \SI{1.5}{\percent} of the total intensity.
It exhibits a broad peak at about \SI{1.8}{\GeVcc} with a bump in its high-mass shoulder. The PDG~\cite{Tanabashi2018} lists the established resonances $K_2(1770)$ and $K_2(1820)$ in the mass region of the peak and the $K_2(2250)$, which needs further confirmation, in the mass region of the high-mass bump.

The $2^-0^+ K^*(892) \pi F$ wave shows a narrow peak at about \SI{1.8}{\GeVcc} with a bump in its high-mass shoulder.
Also the $2^-0^+ \rho(770) K F$ wave exhibits a peak at about \SI{1.8}{\GeVcc}. Its high-mass shoulder contains more complicated structures, which potentially also contain the $K_2(2250)$ resonance.
To clarify the nature of the observed signals and to extract their parameters, i.e. masses and widths, a resonance-model fit similar to Ref.~\citenum{Akhunzyanov2018} will be the next major step in this analysis.

%=============================================================================
%=============================================================================
%=============================================================================
%\section{Conclusion}
%COMPASS has collected the so-far world largest data set of diffractively produced \Kpipi final state of about \num{720000}.
%We performed an elaborate partial-wave analysis, using model selection techniques to select the wave set from a large systematically constructed pool of allowed partial waves.
%The partial-wave decomposition reveals clear signals in the mass region of well known states. However, we observe systematic effects in some partial waves, e.g. in the $1^+0^+K^*(892)\pi S$ wave, from the limited kinematic range of our final-state particle identification.
%In addition, we observe potential signals from excited states.
%To clarify the nature of the observed signals and to extract their parameters, i.e. masses and widths, a resonance model fit is the next major step in this analysis.

\bibliographystyle{ws-procs9x6} % for numbered citation & references
\bibliography{ws-pro-sample}

\begin{thebibliography}{1}

\bibitem{Adolph2015}
C.~Adolph {\em et~al.}, {Resonance production and $\pi\pi$ $S$-wave in $\pi^- +
  p \to \pi^-\pi^-\pi^+ + p_\text{recoil}$ at $190\,\text{GeV}/c$}, {\em
  Physical Review D} {\bf 95}, p. 032004  (2017).

\bibitem{Akhunzyanov2018}
M.~Aghasyan {\em et~al.}, {Light isovector resonances in $\pi^- p \to
  \pi^-\pi^-\pi^+ p$ at 190 GeV/$c$}, {\em Physical Review D} {\bf 98}, p.
  092003  (2018).

\bibitem{kaspar2019}
F.~Kaspar {\em et~al.}, Wave-selection techniques for partial-wave analysis in
  light-meson spectroscopy, within these proceedings.

\bibitem{Daum1981a}
C.~Daum {\em et~al.}, {Diffractive production of strange mesons at 63 GeV},
  {\em Nuclear Physics B} {\bf 187}, 1  (1981).

\bibitem{Tanabashi2018}
M.~Tanabashi {\em et~al.}, {Review of Particle Physics}, {\em Physical Review
  D} {\bf 98}, p. 030001  (2018).

\end{thebibliography}

\end{document}